\documentclass[preprint,journal,hideappendix]{vgtc}        

\vgtccategory{Research}

\title{Trust Your Gut: Comparing Human and Machine Inference from Noisy Visualizations}

\author{
\authororcid{Ratanond Koonchanok}{0000-0002-8860-6183}, \authororcid{Michael E. Papka}{0000-0002-6418-5767}, \authororcid{Khairi Reda}{0000-0002-8096-658X}
}

\authorfooter{
R. Koonchanok and K. Reda are with Indiana University Indianapolis. Email: \{rkoonch, redak\}@iu.edu. M. E. Papka is with Argonne National Laboratory and the University of Illinois Chicago. Email: papka@anl.gov
}

\abstract
{People commonly utilize visualizations not only to examine a given dataset, but also to draw generalizable conclusions about the underlying models or phenomena. Prior research has compared human visual inference to that of an optimal Bayesian agent, with deviations from rational analysis viewed as problematic. However, human reliance on non-normative heuristics may prove advantageous in certain circumstances. We investigate scenarios where human intuition might surpass idealized statistical rationality. In two experiments, we examine individuals' accuracy in characterizing the parameters of known data-generating models from bivariate visualizations. Our findings indicate that, although participants generally exhibited lower accuracy compared to statistical models, they frequently outperformed Bayesian agents, particularly when faced with extreme samples. Participants appeared to rely on their internal models to filter out noisy visualizations, thus improving their resilience against spurious data. However, participants displayed overconfidence and struggled with uncertainty estimation. They also exhibited higher variance than statistical machines. Our findings suggest that analyst gut reactions to visualizations may provide an advantage, even when departing from rationality. These results carry implications for designing visual analytics tools, offering new perspectives on how to integrate statistical models and analyst intuition for improved inference and decision-making. The data and materials for this paper are available at \url{https://osf.io/qmfv6}

}


\graphicspath{{figs/}{figures/}{pictures/}{images/}{./}} 

\usepackage{tabu}                      
\usepackage{booktabs}                  
\usepackage{lipsum}                    
\usepackage{mwe}                       

\usepackage{mathptmx}                  
\usepackage{amsmath}
\usepackage{soul}
\usepackage{balance}

\renewcommand\hl[1]{#1}

\begin{document}


\firstsection{Introduction}

\maketitle
\firstsection{Introduction}

Visualizations play an increasingly important role in data analysis and decision-making. Crucially, these tools are not just used for extracting numbers; often, their greater value lies in revealing insights about the processes that \emph{generated}
the data in the first place. \hl{For example, a business analyst who observes a visualization showing an increase in luxury coat sales might seek to understand the underlying factors, such as unusually cold winters, an increase in disposable incomes, or whether this trend was simply a fluke.} This deeper understanding can in turn help analysts perform higher-level tasks, such as attributing causality, validating hypotheses, or predicting future observations. Making accurate inferences from visualizations, however, can be challenging, requiring one to account for potential uncertainties and natural variabilities in the data. This interpretive process could expose the viewer to pitfalls. For example, the viewer might overinterpret a visualization displaying an unusual or extreme sample, causing them to infer spurious features not found in the data-generating process. Conversely, the viewer may fail to consider the data sufficiently and instead fall back to their prior belief --- a potential manifestation of confirmation bias. Inferential errors during visual analysis can lead to invalid conclusions and false discoveries~\cite{zgraggen2018investigating}. While the likelihood of incorrect machine inference can be quantified analytically or through simulations, our understanding of human inference-making from visualizations remains limited.

Prior research has approached this question by comparing the visual inferences people make to those of an optimal Bayesian agent that observes the same data~\cite{kim2019bayesian}. This method evaluates whether a viewer's interpretation, such as the level of correlation between two variables, aligns with the conclusions of a Bayesian that accounts for both the viewer's preexisting knowledge and the data observed~\cite{karduni2020bayesian}.  The level of divergence between the viewer's inference and the Bayesian model determines how optimal the viewer's interpretation is; visualization techniques eliciting smaller divergence are deemed more effective. 

Although useful as a reference point, the suitability of a fully normative agent as a universal benchmark target is not always justified on an ecological level~\cite{crupi2023critique}. Unlike statistical machines, humans do not possess the necessary cognitive resources to process information like a perfectly rational agent. Additionally, rational agents do not exhibit the nuances of human cognition. In certain contexts, deviations from rationality can indeed be advantageous, leading to a closer approximation of the underlying reality. For example, a Bayesian agent might perceive a positive correlation between two unrelated variables, such as the number of ice cream cones sold and the number of shark attacks, despite a prior suggesting no connection. In contrast, a human observer could become increasingly skeptical upon observing the same dataset, irrationally reinforcing their belief in no correlation. While the Bayesian approach is normatively superior, the human response could lead to better inference in such scenarios. Specifically, the non-rational response of reinforcing a `null' belief in the face of contrary evidence can instill healthy skepticism, guarding against false positives.

In this work, we move away from the assumption that a rational agent is always ideal. Instead, we compare the \emph{utility} of both human and Bayesian inference in accurately characterizing data-generating processes
under varying uncertainty and sample conditions. Specifically, we investigate situations where human visual inferences might outperform those of an ideal Bayesian (and vice versa). Although we expect a Bayesian agent to be globally optimal, we foresee scenarios where factors such as sample size and outlyingness could give the human analyst an edge. This perspective acknowledges that human heuristics, though seemingly inferior~\cite{tversky1974judgment}, can sometimes give rise to better inference~\cite{gigerenzer2009homo}. By characterizing factors that lead analysts to exceed machine-inference performance, we pave the way for designing more intelligent visual analytics systems. Such systems might adaptively encourage analysts to leverage their intuition and `trust their gut', or, at other times, provide computational assistance when users are more prone to biased inference-making. 

We conducted two experiments to study how people make visual inferences in response to bivariate visualizations on a range of topics. Participants externalized both their prior beliefs (i.e., before seeing data) and posteriors (i.e., post-data exposure) using a graphical elicitation device. We manipulated several factors, including the visualization type, sample size, and the extremeness of the sample (i.e., the degree to which it is \emph{inconsistent} with the ground truth). We find that participants were more accurate at inferring the true correlation level than both informed and uninformed Bayesian models when the visualization showed extreme samples. This advantage was particularly pronounced with icon arrays, and in situations where there was a higher consensus around the ground truth. However, participants were less precise in estimating the true uncertainty of the data-generating process, often displaying unwarranted confidence. In a second experiment that varied the uncertainty level in the generating models, participants were better at inferring the true correlation at low uncertainty, although they did not necessarily improve their performance relative to machines.

Our findings reveal that individuals frequently diverge from normative inference when interpreting visualizations. Importantly, these deviations proved advantageous, particularly when examining extreme datasets and small samples. In such instances, an analyst's intuition may yield more accurate conclusions on average than those generated by idealized statistical machines, even when the latter are endowed with human priors.  Conversely, for larger and more reliable datasets, the precision offered by statistical inference seems to reduce both bias and variability, leading to better estimation of the generating model. These findings highlight the potential for integrating human and machine inference in visualization tools, providing guidance on how to interweave the two capabilities. Moreover, our results suggest an alternative design strategy that seeks to harness human heuristics in visual analytics, as opposed to entirely aligning analyst behaviors with rational frameworks. We discuss the implications and highlight future research directions.

\section{Background \& Related Work}

\subsection{Heuristic vs. Rational Decision-Making}

Heuristics are commonly associated with the notion of ``taking shortcuts'' to simplify a cognitively demanding task. As such, heuristics have been viewed as an inferior and unreliable form of cognition compared to rational decision-making~\cite{kahneman1982judgment, gilovich2002heuristics}. However, recent research challenges this perspective, showing the potential of heuristics in helping people come to good decisions~\cite{albrechtsen2009can, gigerenzer2008heuristics}. In certain situations, intuitive decision-making with heuristics can lead to superior outcomes compared to rational, analytical reasoning~\cite{dane2012should, sadler2004intuitive}, especially in high-risk, high-uncertainty scenarios~\cite{huang2018role}. For instance, managers employing heuristics sometimes make more effective decisions than those relying on statistical procedures like logistic regression~\cite{luan2019ecological}. In particular, when information is limited, individuals could achieve better judgments by using simple heuristics rather than attempting to process the available information~\cite{gigerenzer2011heuristic}. In other words, by adaptively disregarding certain information, one could enhance their decision quality~\cite{gigerenzer1999fast}.

Non-rational thinking could also prove beneficial in visual analytics, especially when one is faced with improbable data. For example, analysts could grow skeptical and deviate from normative practices by deliberately underweighting the evidence implied by the data. Such heuristics, which are not always amenable to algorithmic modeling, can enhance inference from visualizations in specific scenarios. Rather than seeking to align analyst performance with normative, statistical inference~\cite{karduni2020bayesian,kim2020bayesian}, we seek a better understanding of how human and machine inference-making can complement each other.

\subsection{Inference from Visualizations}

Statistical inference involves drawing conclusions about a population based on a (limited) sample of data from that population. Although inference is often performed with the aid of formal statistical models, it is possible to make inferences from visualized data~\cite{buja2009statistical}. For visual inferences to be reliable, one must be able to distinguish between real effects and spurious patterns~\cite{zgraggen2018investigating}. Towards that end, methods have been proposed to safeguard the graphical inference process by controlling the rate of false discovery~\cite{zhao2017controlling, savvides2019significance,savvides2022visual}. The lineup protocol is one of the early influential techniques in this space \cite{wickham2010graphical}, and is often considered a visual analog to null-hypothesis significance testing~\cite{majumder2013validation}. Lineups, however, can be difficult to use in practice due to the need to develop realistic null models that can be compared to real datasets~\cite{beecham2016map}.

More recent work aims to allow visual analysts to operationalize their prior beliefs in the interpretation of visualizations~\cite{mahajan2022vibe}. This approach provides affordances for analysts to assess the compatibility of their models with data~\cite{kale2023evm,koonchanok2021data,choi2019visual,choi2019concept,reda2016modeling}. By encouraging analysts to explicitly test their models, researchers aim to foster a more normative, Bayesian-grounded inference from visualizations~\cite{kim2019bayesian,karduni2020bayesian}. Generally speaking, Bayesian inference provides a way to tap into expert knowledge~\cite {dane2012should}, encoding the latter as prior models that can help normalize against outlying data, or supply additional information when data is limited. For example, in climate modeling, prior information can improve estimates of rainfall in the face of potentially extreme measurements~\cite{coles1996bayesian, coles1996bayesian2}. Similarly, in wildlife research, where species might not have the same observable abundance, borrowing information from related species can improve inference about rare populations~\cite{mackenzie2005improving}. Koonchanok et al. propose a parallel mechanism in visualizations~\cite{koonchanok2023visual}. They demonstrate that eliciting priors can reduce false discoveries by allowing analysts to remain vigilant against spurious visualizations. Karduni et al. investigated how individuals update their beliefs when making sense of bivariate visualizations~\cite{karduni2020bayesian}. They show that incorporating uncertainty representations enables a more Bayesian-like belief update. Others have explored how graphical model elicitation might engender attitude change through increased cognitive processing of visualizations~\cite{rogha2024impact,heyer2020pushing,markant2023data}. Building on prior work in belief elicitation~\cite{mahajan2022vibe,koonchanok2021data,koonchanok2023visual,choi2019concept}, we employ interactions to externalize people's priors and posteriors, and compare human performance to that of Bayesian agents. 

\section{Research Questions \& Methods}

We investigate the accuracy of human inference-making from visualizations, evaluating how well these inferences align with the data-generating processes under various uncertainty and visualization conditions. Our central hypothesis is that although analysts will deviate from Bayesian rationality (and thus make suboptimal inferences in the aggregate), they can still update their beliefs in ways that more accurately reflect reality. Specifically, non-normative belief updating, guided by intuition and hunches, could be useful in high uncertainty or noisy sample conditions. In effect, human inference-making can complement strictly rational agents. Thus, we pose two questions:

\vspace{.5\baselineskip}\noindent\textbf{RQ1: } How do visual analysts compare to Bayesian agents that see that same data? How well do humans perform under different visualizations? We compare visual inferences of analysts to two Bayesians: one informed with the prior knowledge of the human analyst, and another with a uniform, flat prior that provides no preexisting knowledge, forcing a purely data-driven inference.

\vspace{.5\baselineskip}\noindent\textbf{RQ2: }How do sample and ground-truth characteristics influence visual inferences relative to Bayesian agents? Factors such as the perceived reliability (or extremeness) of the sample and the degree of ground consensus behind the generating process may impact how individuals derive insights about the data-generating mechanism.

\vspace{.5\baselineskip}\noindent To address these questions, we conducted two crowdsourced experiments. We evaluate individuals' accuracy in inferring true bivariate relationships between attribute pairs after exposure to (potentially noisy) samples. We collected responses from viewers before and after exposure to visualizations, recording both their prior beliefs and posterior inferences about the data-generating process. We measure the accuracy of human-vs-Bayesian inferences under different sample configurations, ranging from large samples that reflect the true parameters to small and potentially extreme samples. Additionally, we assess the effectiveness of multiple visualization types for this purpose.

\subsection{Model Elicitation}
\label{sec:elicitation}

\begin{figure}[t]
\center
\includegraphics[width=1\linewidth]
{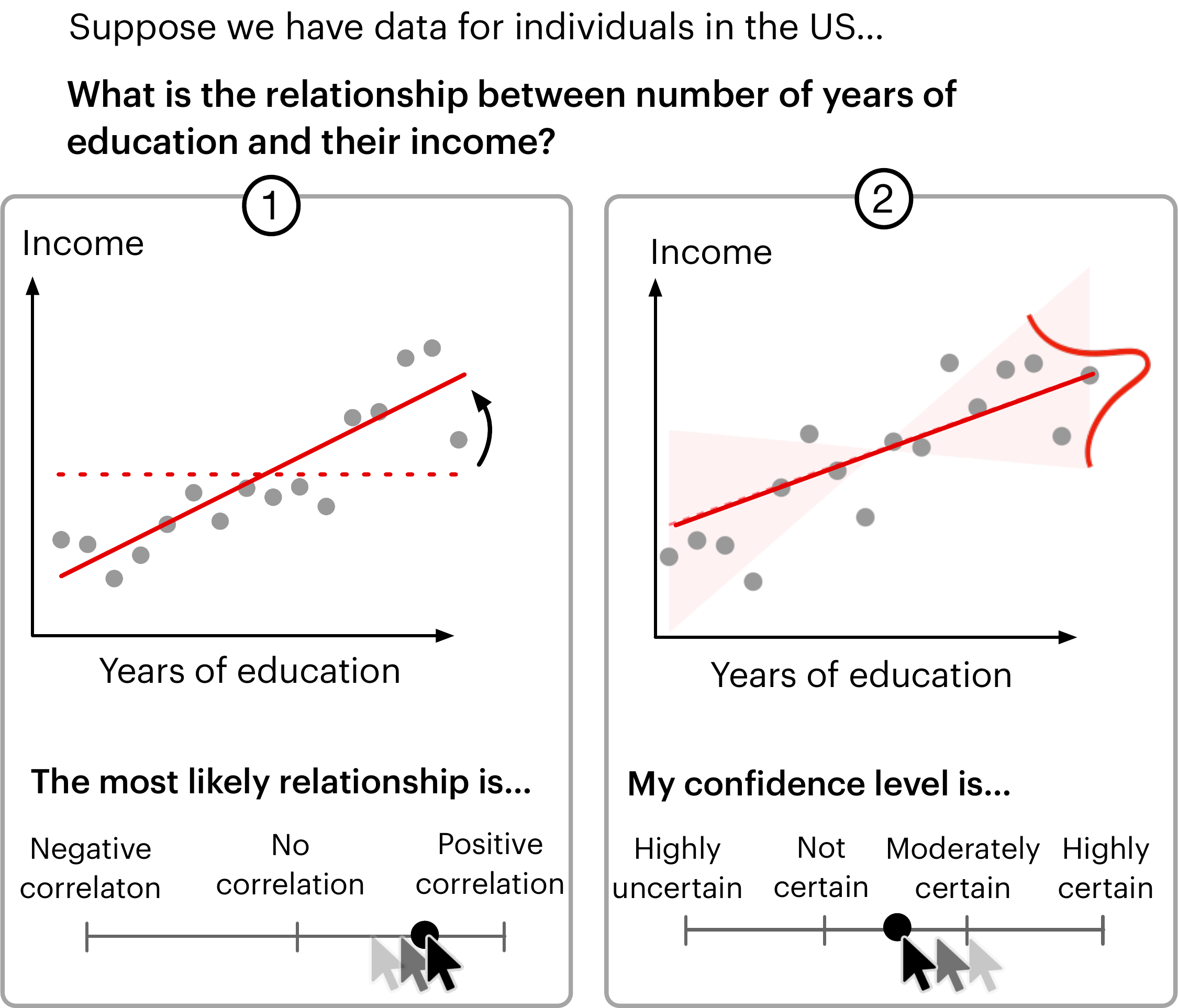}
  \caption{
  A graphical belief elicitation device for expressing beliefs about bivariate relationships in response to a prompt (top). \hl{Participants externalized their prior and posterior beliefs in two steps: (1) indicating the most likely relationship, and (2) specifying their uncertainty in the relationship. These slider settings update two parameters, $\mu$ and $\sigma$, in a linear model. During this interaction, participants see hypothetical samples from this model, refreshed at 5Hz, illustrating what the bivariate data might look like if their beliefs were true.
  }}
  \label{fig:elicitation}
  \vspace{-5mm}
\end{figure}

We employ a graphical elicitation device to externalize participants' prior and posterior beliefs (illustrated in Figure \ref{fig:elicitation}). This interface enables participants to specify two parameters via two sliders according to their beliefs: the expected correlation coefficient between the two variables ($\mu$) and the associated uncertainty in the correlation ($\sigma$). Specifically, the first slider prompts participants to indicate the ``most likely relationship'' along a continuum from `negative' to `positive' correlation. The second slider prompts participants to express their ``confidence level'' in the relationship, ranging from `highly uncertain' to `moderately' and `highly certain'. These two parameters  populate the following model: 

\begin{equation} 
\begin{aligned}
    y_i = \beta_0 + \beta  x_i + \epsilon_i \\ \label{eq:model}
    \beta \sim \mathcal{N}(\mu, \sigma^2)\\
    \beta_0 \sim \mathcal{N}(0, \sigma_b^2) \\
    \epsilon_i \sim \mathcal{N}(0, {\sigma_e^2})
\end{aligned}
\end{equation}

Where $\mu$ is the \emph{expected slope} of the relationship as specified by the first slider, and $\sigma$ is the \emph{uncertainty} in the slope, as specified through the second slider. The intercept $\beta_0$ for the regression line is centered around 0, with a fixed standard deviation of $\sigma_b=0.1$. $\epsilon_i$ is an additional residual term with a fixed standard deviation of $\sigma_e=0.45$.

During user interaction, the elicitation interface visualizes samples generated from the above model. \hl{This allows participants to see hypothetical outcomes, refreshed at 5Hz, showing what the data \emph{might} look like if their belief was true. This type of animated display is meant to help participants intuitively grasp the effects of manipulating the two sliders and the uncertainty implied by their (prior) model}. \hl{The hypothetical samples are displayed in a scatterplot} (Figure~\ref{fig:elicitation}), \hl{a bar chart, or an icon array} (Figure~\ref{fig:steps}-A and B), \hl{depending on the experimental condition. However, in all conditions, participants utilize the same two sliders to specify their beliefs. The hypothetical samples are only animated during the elicitation step and shown in grey color to help distinguish them from the ground truth data samples (solid black).}

\subsection{Stimuli and Data-Generating Models}
\label{sec:questions}

\begin{table}[b]
    \vspace{-4mm}
    \centering
    
    \caption{Example prompt questions and associated ground-truth parameters as obtained from crowd workers.}
    \label{tab:example_questions}
    \includegraphics[width=1\linewidth]{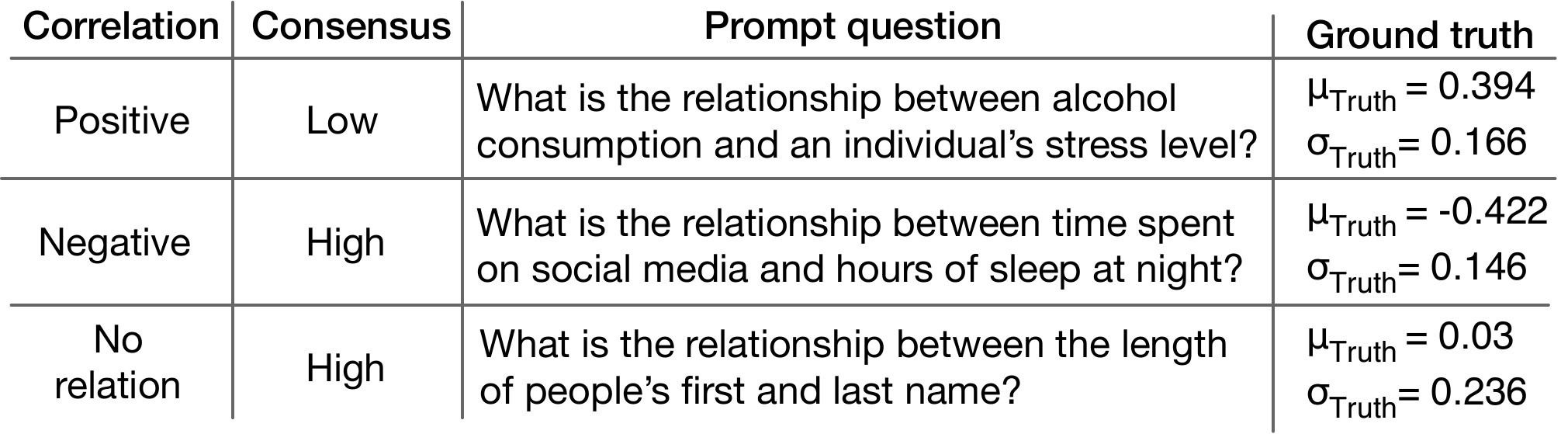}
\end{table}

\begin{figure}[t]
\center
\includegraphics[width=1\linewidth]
{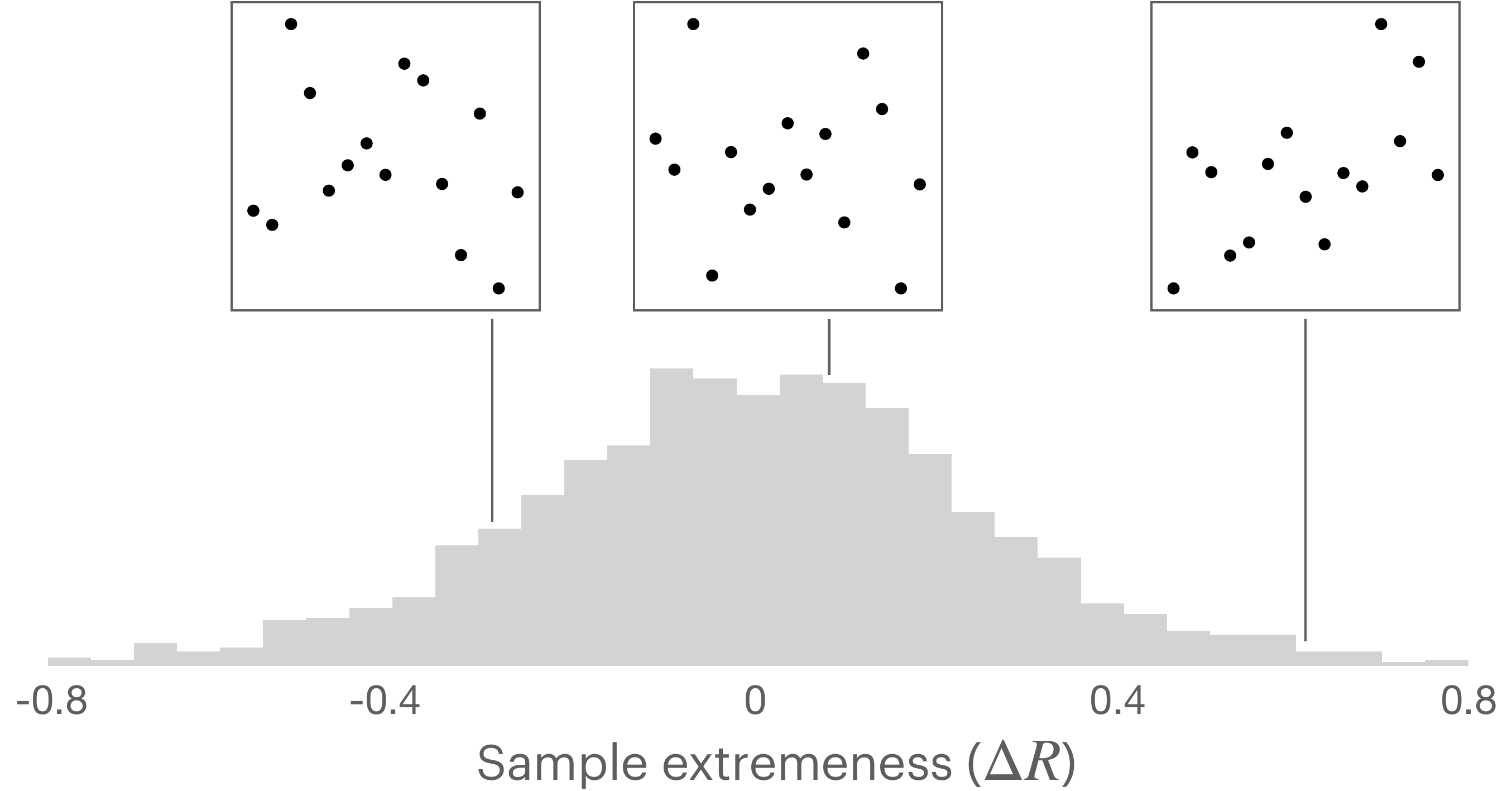}
  \caption{Distribution of extremeness for `medium' sample sizes ($N=15$ points). With $\Delta R \approx 0.1$, the middle sample is a relatively faithful depiction of the ground truth (no correlation in this example).  The sample on the left is more extreme with $\Delta R \approx -0.3$. The right-most scatterplot represents an even more extreme occurrence with $\Delta R \approx 0.6$.}
  \label{fig:extremeSamples}
  \vspace{-4mm}
\end{figure}

To provide plausible stimuli for the study, we developed a set of 40 prompt questions and corresponding ground truth models. The prompt questions covered a variety of common knowledge topics (see Table~\ref{tab:example_questions} for examples), with bivariate relationships ranging from negative correlations to positive correlations. Additionally, we also included attributes with no plausible relationship. The corresponding ground truth models for these prompts followed the same form as Equation~\ref{eq:model}. To initialize these models with plausible parameters rooted in common wisdom, we employed crowd workers recruited through Amazon Mechanical Turk. Workers ($n=61$) were prompted to respond to each of the 40 questions, using the model elicitation interface in Figure~\ref{fig:elicitation} to provide their belief on the most likely relationship slope and their uncertainty around that relationship.  Responses from the workers were then averaged forming the two ground truth parameters for each prompt question. Subsequently, we selected 24 prompt questions to be used in the experiment. These questions comprised six ground truth models with a positive relationship ($\mu_{Truth}>0$), six questions with a negative relationship ($\mu_{Truth}<0$), and 12 with no correlation ($\mu_{Truth}\approx0$).

\vspace{.7\baselineskip}\noindent\textbf{Social Consensus: } Individuals often rely on social knowledge when forming their beliefs. Consensus (whether perceived or actual) can also serve as a tool to reduce uncertainty. Therefore, we expect the agreement around the ground truth to impact people's inferences from visualizations. To quantify the latter, we measure the consistency of the crowd wisdom: Prompt questions exhibiting smaller variations in the elicited crowd beliefs are considered to reflect a higher degree of social consensus. This was determined based on the standard deviation of the elicited $\mu$ responses among workers. Within each category (positive, negative, or no relationship), we designate half the prompt question with the lowest standard deviation as `high' consensus, with the other half considered `low' consensus, representing lower agreement between workers on what the data-generating process should be. \hl{Including this factor in our analysis helps control for prior beliefs. Participants are expected to have stronger priors when responding to high-consensus questions. Conversely, priors are likely to be much weaker for low-consensus questions, leading participants to become more data-driven in their inference, and potentially more susceptible to extreme samples.}

\subsection{Sample Configurations}
\label{sec:sampleProperties}

For each stimulus, we display the prompt question and ask the participant to provide their prior belief about the topic ($\mu$ and $\sigma$). We then expose participants to a random data sample generated from the ground truth model. Subsequently, we prompt them to reflect and provide a posterior inference using the same elicitation device as before. To understand how sample characteristics affect inference accuracy, we varied two properties: sample size and extremeness. 

\vspace{.7\baselineskip}\noindent\textbf{Sample Size: }The number of data points in a sample is an important factor for an analyst to consider when making inferences. Larger samples furnish stronger evidence about the underlying data-generating process. We thus varied the size of samples shown to participants, using 7, 15, and 30 data points to represent `small', `medium', and `large' sample sizes, respectively (\hl{these labels are only to illustrate the results, and not shown in the experiment}). These sizes were selected to provide varying levels of evidence, while still allowing for extreme samples to emerge. Specifically, smaller samples are more prone to noise, giving a potentially misleading picture of the underlying ground truth. 

\vspace{.7\baselineskip}\noindent\textbf{Sample Extremeness: }A core question that we address is how resilient people can be to extreme (or spurious) data, as compared with idealized statistical machines. Sample \emph{extremeness} reflects the degree to which it is \emph{inconsistent} with the data-generating process (e.g., a sample suggesting a positive correlation when the underlying model prescribes no relationship). Under a random sampling regime, spurious samples are expected to arise. We take advantage of this phenomenon and quantify the extremeness of the emerging samples in the experiment. We use Pearson's coefficient ($R$) to measure correlation strength for a sample and, by extension, the degree to which it can be considered \emph{extreme} with respect to its underlying model. Specifically, we compute a difference ($\Delta R$) between the sample's and the \emph{expected} correlation:

\begin{figure*}[]
\center
\includegraphics[width=1\linewidth]{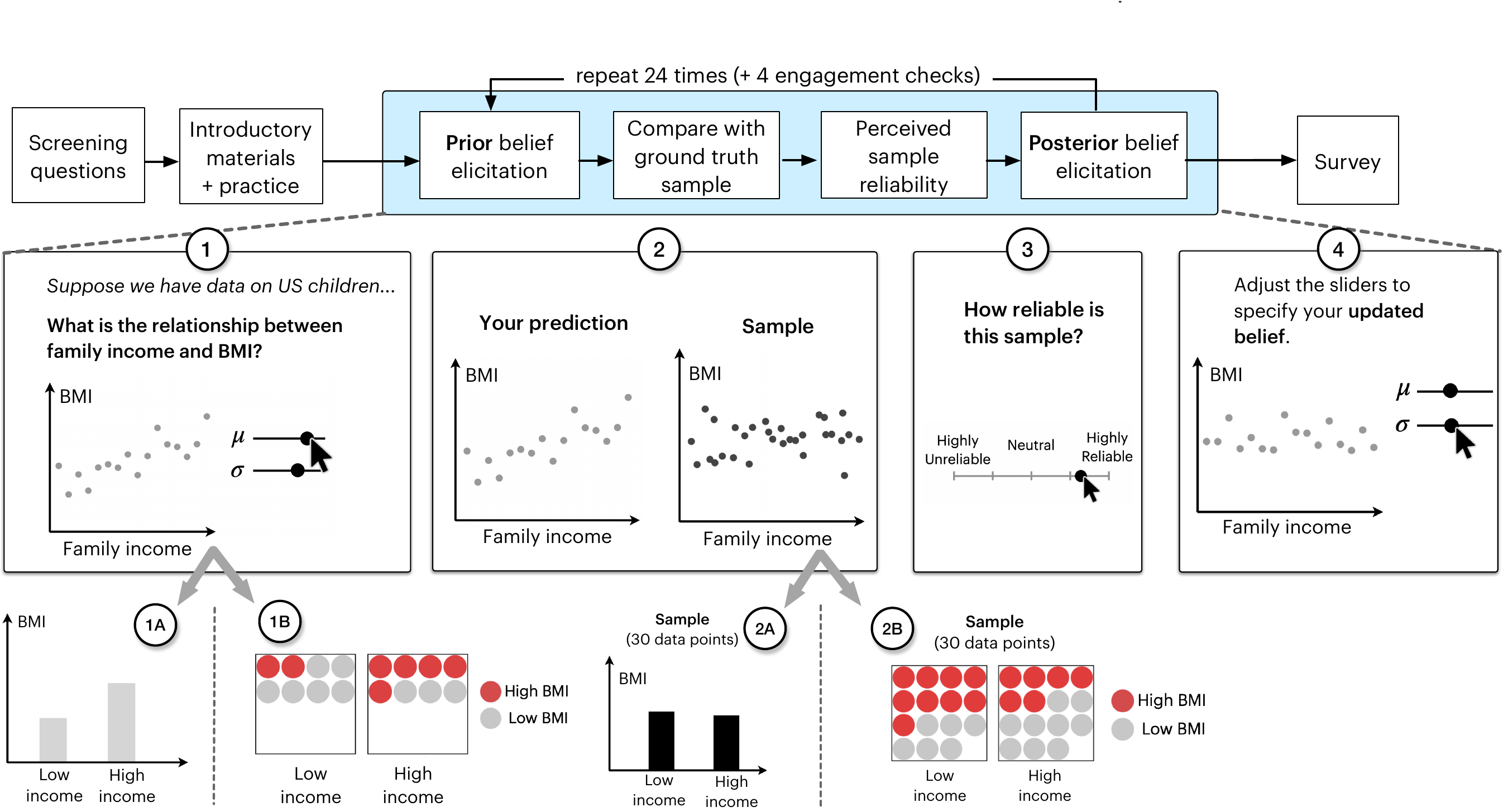}

  \caption{\hl{Top: Flow diagram illustrating the procedure for Exp. 1. The blue-shaded section represents a single trial. In each trial, participants: 1) externalize their prior belief about a prompt question by adjusting two sliders, 2) observe a sample from the ground truth presented alongside their belief, 3) indicate how reliable they believe the sample is, and 4) specify their updated (posterior) belief by re-adjusting the sliders. Depending on the visualization condition, participants are either shown scatterplots, bar charts (A), or icon arrays (B).}}
  
  \label{fig:steps}
  \vspace{-5mm}
\end{figure*}

\begin{equation}
\begin{aligned}
\Delta R = R_{Sample} - R_{Expected} \label{eq:extremeness}
\end{aligned}
\end{equation}

$R_{Expected}$ is the mean R-value, determined from 10,000 simulated draws from the ground truth. A sample that is a faithful representation of its model will have $\Delta R \approx 0$. As $\Delta R$ deviates (negatively or positively), the sample will be considered more extreme. Under the law of large numbers, we would expect the distribution of $\Delta R$ to be centered around zero, with the majority of samples exhibiting low extremeness (see Figure~\ref{fig:extremeSamples}). However, extreme samples would be expected to emerge. Additionally, we anticipate that sample size will influence the likelihood of extremeness: larger samples are expected to have a narrower $\Delta R$ distribution, while smaller samples may yield a wider extremeness distribution. Ground-truth models with higher uncertainty ($\sigma$) will also produce more spurious data. By quantifying the extremeness of emerging samples, we can observe how participants react to spurious data. We can then assess the extent to which participants (and Bayesian agents) can resist misleading samples.
\section{Experiment I}

This experiment aims to compare the inference of participants to ideal Bayesian agents. We specifically test the extent to which participants can infer the ground truth parameters as a function of visualization type, sample size and extremeness, and the degree of social consensus. Participants saw prompt questions described in \S\ref{sec:questions} and were asked to predict the parameters of a bivariate ground truth model, namely, the expected slope $\mu$ and uncertainty $\sigma$. For each stimulus, participants were asked to provide their prior beliefs through the graphical elicitation interface described earlier. After that, they saw a sample from the model and were asked to reflect and provide their updated (i.e., posterior) inference about the true relationship.

\subsection{Experiment Design}

We adopted a mixed design, investigating three independent factors: Visualization (3 types) $\times$ Sample size (3 levels) $\times$ Social consensus (2 levels). Visualization was varied between subjects, whereas sample size and social consensus were varied within-subject.

\vspace{.5\baselineskip}\noindent \textbf{Visualization}: We evaluated scatterplots, bar charts, and icon arrays for representing samples. Scatterplots were chosen for their perceptual effectiveness in illustrating correlations~\cite{harrison2014ranking}. Bar charts on the other hand aggregate the X axis into ordered groups (e.g., `low' and `high income'), averaging the second attribute (e.g., BMI) within these groups (see Figure~\ref{fig:steps}-A). Icon arrays similarly aggregate the X axis into discrete `low' and `high' bins but show individual data points with color-coded Y values (Figure~\ref{fig:steps}-B). While scatterplots offer a full view of the sample, both bar charts and icon arrays provide simplified, aggregate representations. Icon arrays were particularly noted for aiding conditional probability assessment of the sort needed for Bayesian reasoning~\cite{micallef2012assessing}. \hl{They have also been used extensively in communicating uncertainty (e.g., medical risks}~\cite{galesic2009using,garcia2010icon}) \hl{to non-experts, making them a suitable visualization to test in this experiment.} That said, the collapse of continuous variables into discrete ordinals in both icon arrays and bar charts may ultimately prove detrimental.

\vspace{.5\baselineskip}\noindent \textbf{Sample and Ground Truth Characteristics}: Participants completed eight trials with each of the three sample sizes (small, medium, and large). Of these eight trials, four were presented with high-consensus prompts and four were low-consensus. Sample extremeness was included as an explanatory variable but was left to vary as a consequence of the random sampling process. Hence, while not explicitly controlled, extremeness covaried following a predictable distribution (Figure~\ref{fig:extremeSamples}).

\subsection{Hypotheses}

We developed four hypotheses:

\vspace{.3\baselineskip}\noindent\textbf{H1 --} Compared to the two Bayesian agents,  
participants will be less accurate \emph{overall} in recovering the true slope of the data-generating process. The global suboptimality of human analysts can be attributed to the difficulty in performing a fully normative inference from data on a visual basis. That said, we hypothesize that the informed Bayesian will outperform the uninformed (i.e., flat prior) agent, as the former benefits from participants' knowledge.

\vspace{.3\baselineskip}\noindent\textbf{H2 --} Although participants will make less-optimal inferences in the aggregate, we expect them to handle extreme samples better than Bayesian agents. Specifically, we hypothesize that participants will employ non-statistically normative heuristics to `adjust' for improbable data, leading to more accurate inference.  We expect this tendency to be more evident when there is greater consensus around the ground truth and with smaller sample sizes.

\vspace{.3\baselineskip}\noindent\textbf{H3 --} Participants will perform best with scatterplots, thanks to their effectiveness in displaying bivariate correlations. Icon arrays should also yield good results, though to a lesser degree. Both visualizations allow for individual data point observation. In contrast, participants will struggle with bar charts due to the absence of uncertainty representations, making this visualization somewhat less optimal for inference.

\vspace{.3\baselineskip}\noindent\textbf{H4 --} We anticipate reduced accuracy in participants' judgment of the model's uncertainty. Specifically, we expect human analysts to perform less effectively than both the informed and uninformed Bayesian agents when characterizing uncertainty. 

\subsection{Participants}
We recruited participants from Prolific who are US residents and are at least 18 years old. Prospective participants were initially screened through basic questions on bivariate relationships to ensure sufficient background. We ultimately enrolled 222 participants in the experiment (112 males, 104 females, 6 others) with a mean age of 35.4 years. \hl{Education levels for participants were as follows: 56 high school, 24 associate, 89 bachelor's, 39 master's, 13 doctorate, and 1 unspecified.} Participants were randomly assigned to one of the three visualization types (74 in each group). They were compensated with \$5 for an hourly wage of \$16.7. The study was approved by Indiana University's IRB.

\subsection{Procedure}

Participants first completed a tutorial explaining the goals of the experiment and introducing the graphical elicitation device. The instructions emphasized that the samples shown could be noisy, particularly for small samples. Participants then completed one practice trial, after which they completed the main study which consisted of 24 trials, corresponding to the prompt questions developed in \S\ref{sec:questions}, presented in random order. In addition to the analyzed trails, we included four engagement checks. The checks instructed participants to set the sliders to specific values (e.g., extreme positive correlation). 

In each trial, participants were first presented with a context (e.g., ``supposed we have data on US cities''), and were then prompted to predict a linear relationship between two variables (e.g., ``what is the relationship between the unemployment rate and the affordability of housing?'') Participants were asked to visually provide their prior belief through the elicitation device (\S\ref{sec:elicitation}). Next, the participants were presented with a sample drawn from the corresponding ground-truth model. The sample was shown side-by-side with participants' prior. They were prompted to provide their impression of sample reliability via a slider. Lastly, participants were prompted to update their beliefs by re-adjusting. \hl{Participants, however, were instructed that it is up to them to decide ``how much (or little) to adjust [their] initial beliefs.''}  Figure~\ref{fig:steps} illustrates this sequence. The visualized samples along with the belief elicitation device varied depending on the visualization condition (scatterplots, bar charts, or icon arrays). However, participants entered their responses using an identical set of sliders, with the difference being in how the samples, priors, and posteriors were visualized.

\subsection{Response and Accuracy Metrics} Participants supplied five continuous responses in every trial: the most-likely slope ($\mu_{Human}$) and uncertainty ($\sigma_{Human}$) in the relationship, both before (i.e., prior belief) and after sample exposure (i.e., posterior inference). Additionally, they provided their perception of sample reliability (reported in the supplemental materials). For every trial, we generated two comparable statistical inferences. The first ($\mu_{Bayes}$ and $\sigma_{Bayes}$) was derived through Bayesian inference using the participant-provided prior and the likelihood function implied by the sample observed by the participant. This corresponds to an idealized statistical agent equipped with identical prior knowledge as the participant. The second response, also a Bayesian, utilized a flat prior in conjunction with the same likelihood function, representing an uninformed agent that exclusively learns the two parameters ($\mu_{Flat}$ and $\sigma_{Flat}$) from the sample. We quantify inference accuracy for each of the three agents by measuring the divergence of the inferred parameters from the ground truth:

\begin{equation}
\begin{aligned} 
   \Delta \mu &= \frac{ \mu_{Human|Bayes|Flat} - \mu_{Truth} }{2} \times 100  \\
    \Delta \sigma &= \frac{ \sigma_{Human|Bayes|Flat} - \sigma_{Truth} }{0.6} \times 100   \\
\end{aligned}
\end{equation}

We normalize relative to the most likely parameter range of $\mu \in [-1, 1], \sigma \in [0, 0.6]$ and report 100$\times$ values for readability.

\subsection{Analysis and Modeling}
\label{sec:bayes_model}

Participants completed the experiment in 18 minutes on average (\emph{sd} = 8.7). They provided 5,328 responses in total. We used the \texttt{brms} package~\cite{burkner2017brms} to fit the responses to a Bayesian regression model. The model predicts  $\Delta \mu$ or $\Delta \sigma$ (i.e., the divergence from true slope or uncertainty). We modeled both the \emph{mean} and \emph{spread (sd)} of the divergence. We followed a Bayesian analysis workflow in creating the model~\cite{gelman2020bayesian}, evaluating with posterior predictive checks and incorporating additional parameters to improve the posterior fit. Details on the model construction process can be found in the supplementary material. The model for $\Delta \mu$ is given by (with a nearly identical formula to predict $\Delta \sigma$):

\begin{equation*}
\begin{aligned} 
    \Delta \mu  &\sim Normal(\text{mean}, \text{sd}^2) \\
    \text{mean} &= \Delta R \times agent \times vis \times consensus \times size \\ 
        & + \Delta R\times agent \times questionType  \\
        & + (1+\Delta R \times agent~|~participant) 
         + (1~|~question) \\
    log(\text{sd}) &= agent \times vis \times consensus \times size \\
        & + (1+size \times agent~|~participant) + (1~|~question) \\
\end{aligned}
\end{equation*}

$\Delta R$ is the sample extremeness (Equation~\ref{eq:extremeness}). \emph{Agent} indicates whether the inference came from a human (the participant), an informed Bayesian, or an uninformed flat-prior agent. \emph{Size} is a categorical variable representing the sample size (small, medium, or large). \emph{Vis} represents the visualization type (scatterplot, bar chart, or icon array). \emph{Consensus} is the agreement around the ground truth. The first interaction term allows us to model our primary effects of interest. After conducting posterior predictive checks, we introduced an effect for \emph{questionType}; this categorical variable signifies whether the ground truth prescribes a positive correlation, negative correlation, or no relationship -- the latter played a role in modulating participants' responses. Lastly, we included random slopes and intercepts to model individual differences among participants and effects of topic variation (\emph{question}). We used flat priors for all parameters, except for the spread which followed a half-student's $T$ distribution ($df=3$, $bias=0$, $scale=2.5$).

\begin{figure*}[t]
\center
\includegraphics[width=1\linewidth]
{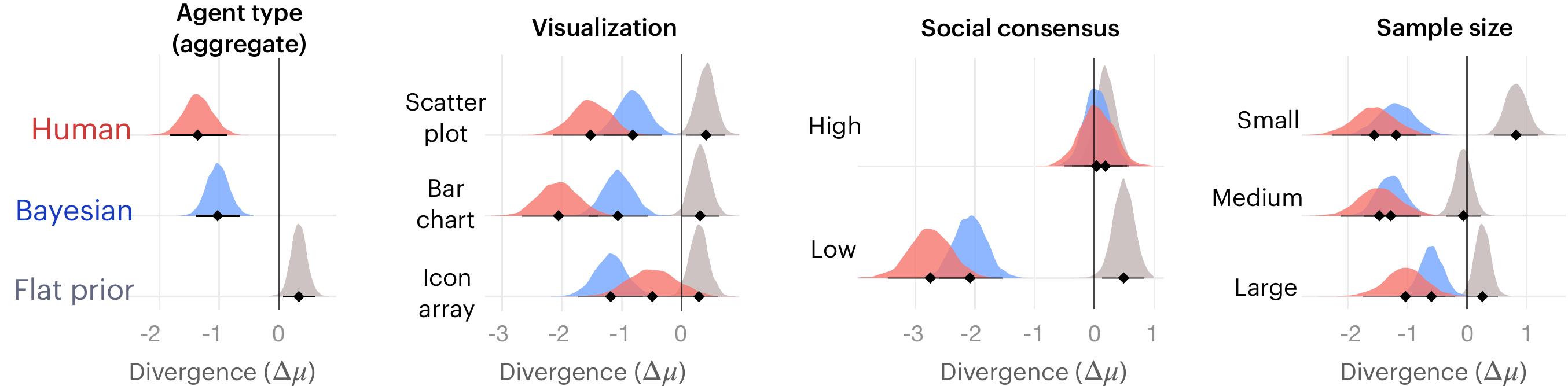}
  \caption{Estimated mean divergence ($\pm$ 95\% credible intervals) for participants vs. informed and uninformed (flat prior) agents. Smaller divergence from zero indicates better accuracy at inferring the true slope. Right: marginalized effects of Visualization, Consensus, and Sample size.}
  \label{fig:exp1_global}
  \vspace{-4mm}
\end{figure*}

\begin{figure}[t]
\center
\includegraphics[width=1\linewidth]
{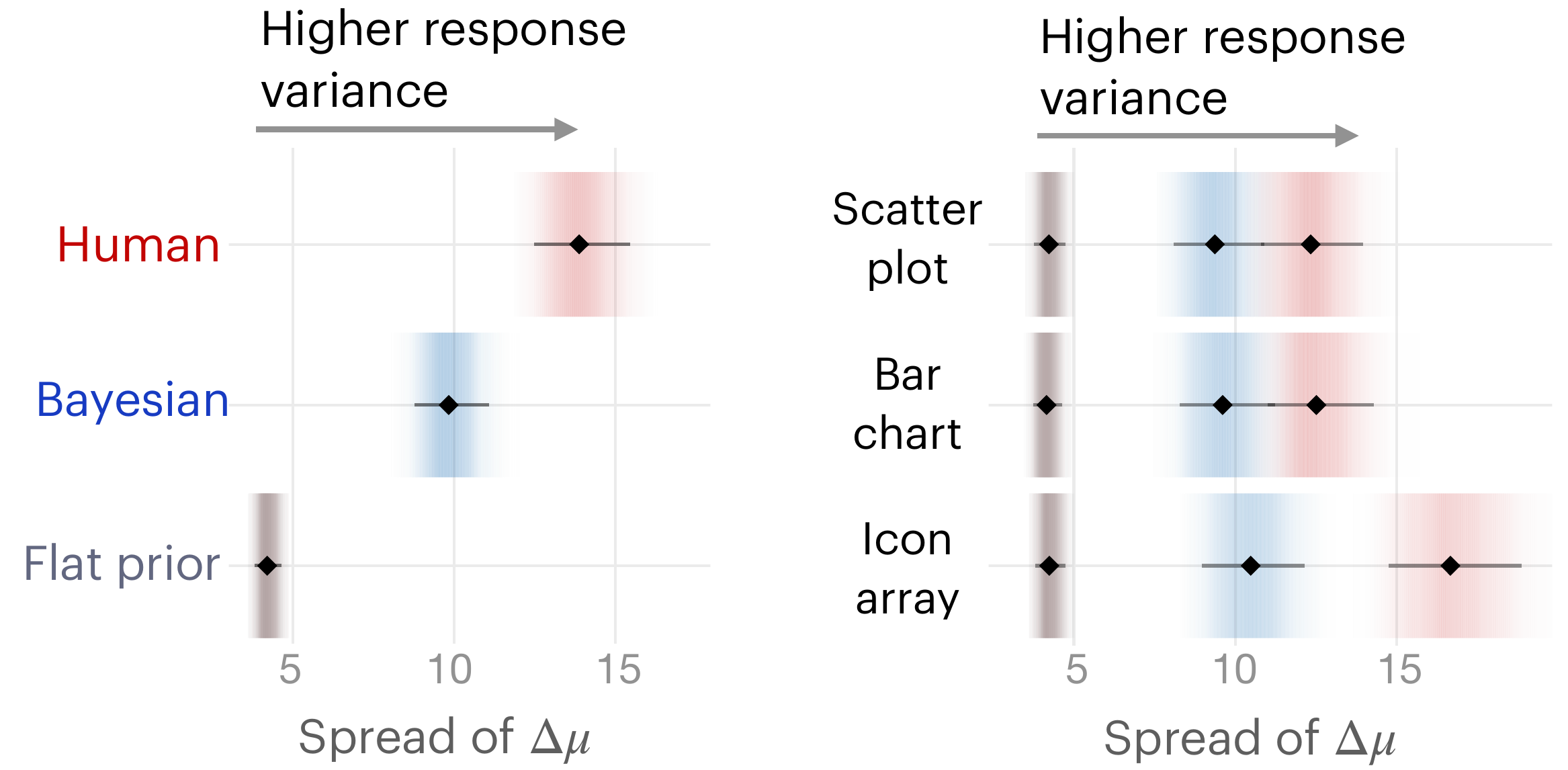}
  \caption{Estimated spread (standard deviation) of $\Delta \mu$ ($\pm$ 95\% CIs) for participants (red) vs. informed and uninformed Bayesians. A larger spread implies higher variability in inference accuracy. The plot on the right shows the marginalized effects of visualization type.}
  \label{fig:exp1_spread}
  \vspace{-3mm}
\end{figure}

\begin{figure}[t]
\center
\includegraphics[width=0.95\linewidth]{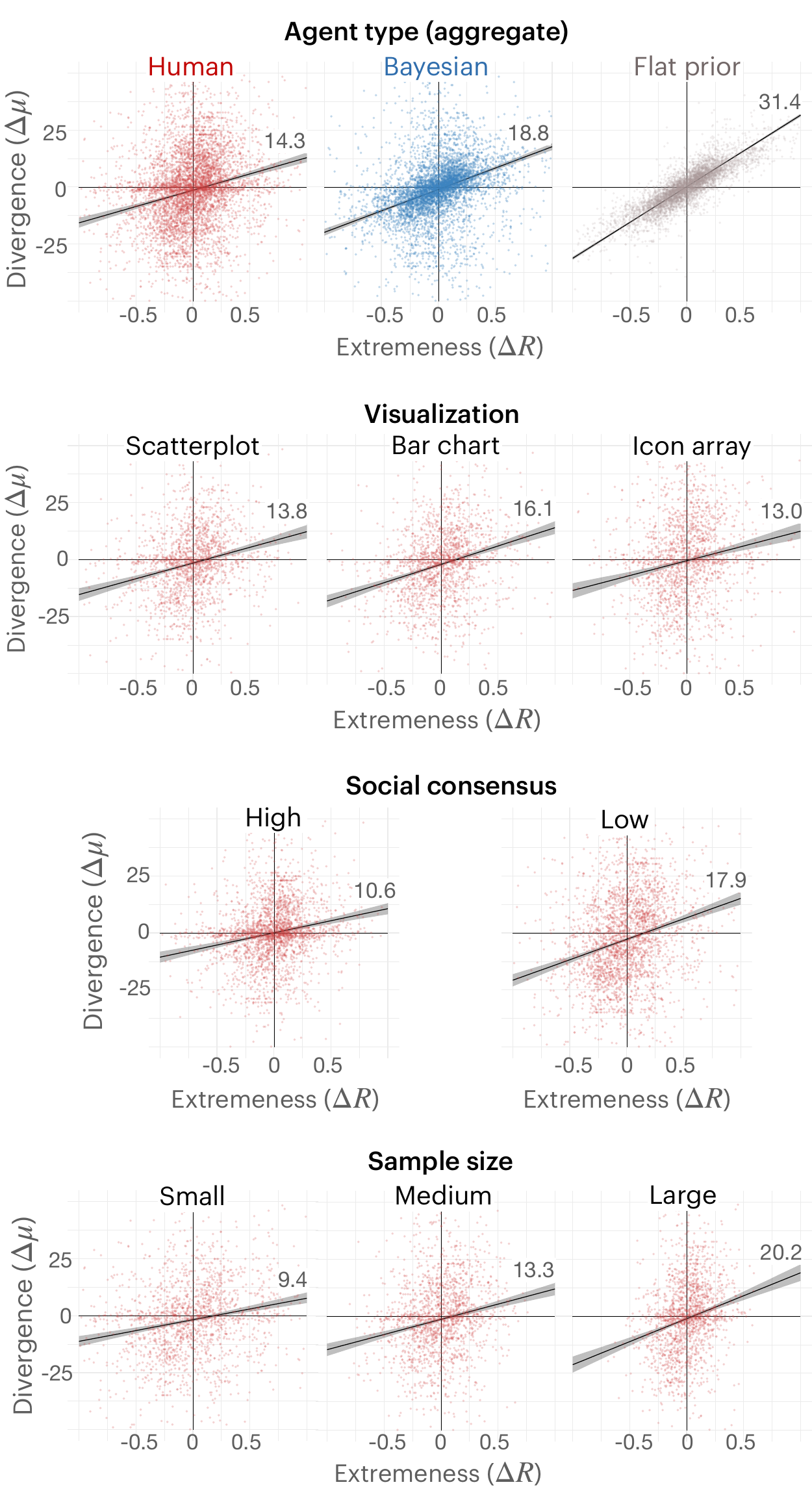}
  \caption{Sensitivity to extreme samples for the three agents (top) and for participants alone (bottom rows). Lines depict mean estimate of ${\Delta \mu}/{\Delta R}$ slopes ($\pm$ 95\% CIs). A {lower slope indicates better resilience} to spurious samples. The point clouds depict the observed responses. }
  \label{fig:exp1_lineCharts}
  \vspace{-4mm}
\end{figure}

\subsection{Results}

\subsubsection{How accurate are participants in inferring the true slope compared to Bayesian agents?}

Figure~\ref{fig:exp1_global} illustrates the mean accuracy of three agents overall and across all manipulated factors. Zero divergence is ideal. Overall, the flat-prior agent exhibits the smallest distance to the real slope (est. divergence: $0.34$, 95\% credible intervals: [$0.08, 0.61$]). This was followed by the informed Bayesian agent ($-1.02$, 95\% CI: [$-1.38, -.65$]), and lastly, participants who provided the least accurate inferences ($-1.36$, CI: [$-1.81, -.87$]). Humans and informed Bayesians overlapped substantially in their accuracy, although they both consistently underestimated the true relationship. By contrast, an uninformed Bayesian appears to provide near-optimal inference on the aggregate. We analyze how accuracy varies across visualization types and social consensus levels.

\vspace{.5\baselineskip}\noindent\textbf{Visualization}: Participants appear to perform best when viewing icon arrays. On average, their divergence from the true slope was only $-.49$ with a 95\% credible interval that overlaps zero ([$-1.27, 0.3$]), indicating a good chance of inferring the true relationship. Notably, human performance with icon arrays slightly surpassed though remained comparable to that of an informed Bayesian ($-1.18$, CI: [$-1.73, -0.65$]). In contrast, participants tended to make less accurate inferences with scatterplots ($-1.52$, CI: [$-2.15, -0.9$]) and bar charts ($-2.06$, CI: [$-2.66, -1.4$]), with both visualizations leading to an underestimation of the true relationship.

\vspace{.5\baselineskip}\noindent\textbf{Consensus}: All agents performed equally well when the consensus around the ground truth was high, with divergence remarkably centered around zero. By comparison, low-consensus topics elicited less accurate inferences from participants (divergence: $-2.75$, CI: [$-3.46, -2.07$]). This reduction in accuracy was also observed to a lesser degree in the informed Bayesian ($-2.08$, CI: [$-2.6, -1.54$]). As would be expected, a flat-prior agent is virtually unaffected by whether there is ground consensus (0.18, CI: [$-0.17, 0.54$] for high consensus vs. $0.49$, CI: [$0.13, 0.84$] for low).

\vspace{.5\baselineskip}\noindent\textbf{Sample Size}: 
The data-driven agent consistently demonstrated good inference across medium and large sample sizes, with credible intervals that intersected zero. As expected, its performance was reduced in small sample sizes, resulting in a non-zero deviation ($0.82$, CI: [$0.46, 1.19$]). This degree of deviation in smaller datasets was somewhat comparable to that of an informed Bayesian ($-1.19$, CI: [$-1.77, -0.6$]), and to human performance ($-1.55$, CI: [$-2.26, -0.86$]). 

\vspace{.5\baselineskip}\noindent\textbf{Spread:} Figure~\ref{fig:exp1_spread} illustrates the estimated spread in $\Delta \mu$. Increased spread corresponds to higher variability in responses.  Overall, the flat prior agent displays the least spread (est. SD: 4.21, CI [$3.81, 4.65$]), indicating a consistent response. By contrast, the informed Bayesian (9.83, CI: [$8.78, 11.1$]) exhibits higher variability. Participants, at the upper end, demonstrate the greatest spread (13.9, CI: [$12.5, 15.5$]), indicating higher variance in their ability to infer the true slope. Figure \ref{fig:exp1_spread}-right further dissects this by visualization type. Among participants, icon arrays led to the widest spread (16.7, CI: [$14.8, 19$]), consistently surpassing the other two visualizations (12.3, CI: [$10.8, 14$] for scatterplot and 12.5, CI: [$11, 14.3$] for bar charts).

\subsubsection{How resilient are humans and Bayesian agents to spurious samples?}

Figure \ref{fig:exp1_lineCharts}-top illustrates the impact of sample extremeness on the divergence from the ground truth for the three agent types. A weaker correlation (i.e., smaller slope) indicates lower sensitivity and, hence, better resilience to spurious samples. Participants appear to be less influenced by extreme samples than both informed and uninformed (flat prior) Bayesian agents. Specifically, a unit increase in sample extremeness leads to a 14.3 (CI: [$12.2, 16.47$]) increase in participant divergence from the true relationship. By contrast, a Bayesian is influenced more strongly (18.80, CI: [$17.25, 20.27$]), giving rise to higher error with spurious samples. As would be expected, an agent with a flat prior learning purely from the data is influenced the most (31.41, CI: [$30.68, 32.1$]), leading to a strong association between sample quality and inference accuracy. 

\vspace{.5\baselineskip}\noindent\textbf{Visualization, Consensus, and Sample Size:} Icon arrays afford the highest resiliency (est. slope: 12.98, CI: [$9.78, 16.55$]), followed by scatterplots (13.79, CI: [$10.94, 16.64$]) and bar charts  (16.10, CI: [$13.41, 18.88$]).  These slopes, however, overlap indicating largely similar effects for the different visualizations. When there is more consensus around the ground truth, resiliency is consistently better (10.62, CI: [$8.19, 13.04$] vs. 17.95, CI: [$15.26, 20.79$]). Lastly, sample size also played a role, with participants least influenced by small samples (9.43, CI: [$7.24, 11.84$]), followed by medium (13.33, CI: [$10.59, 16.02$]) and large (20.18, CI: [$16.59, 23.74$]).

\begin{figure}[t]
\center
\includegraphics[width=1\linewidth]
{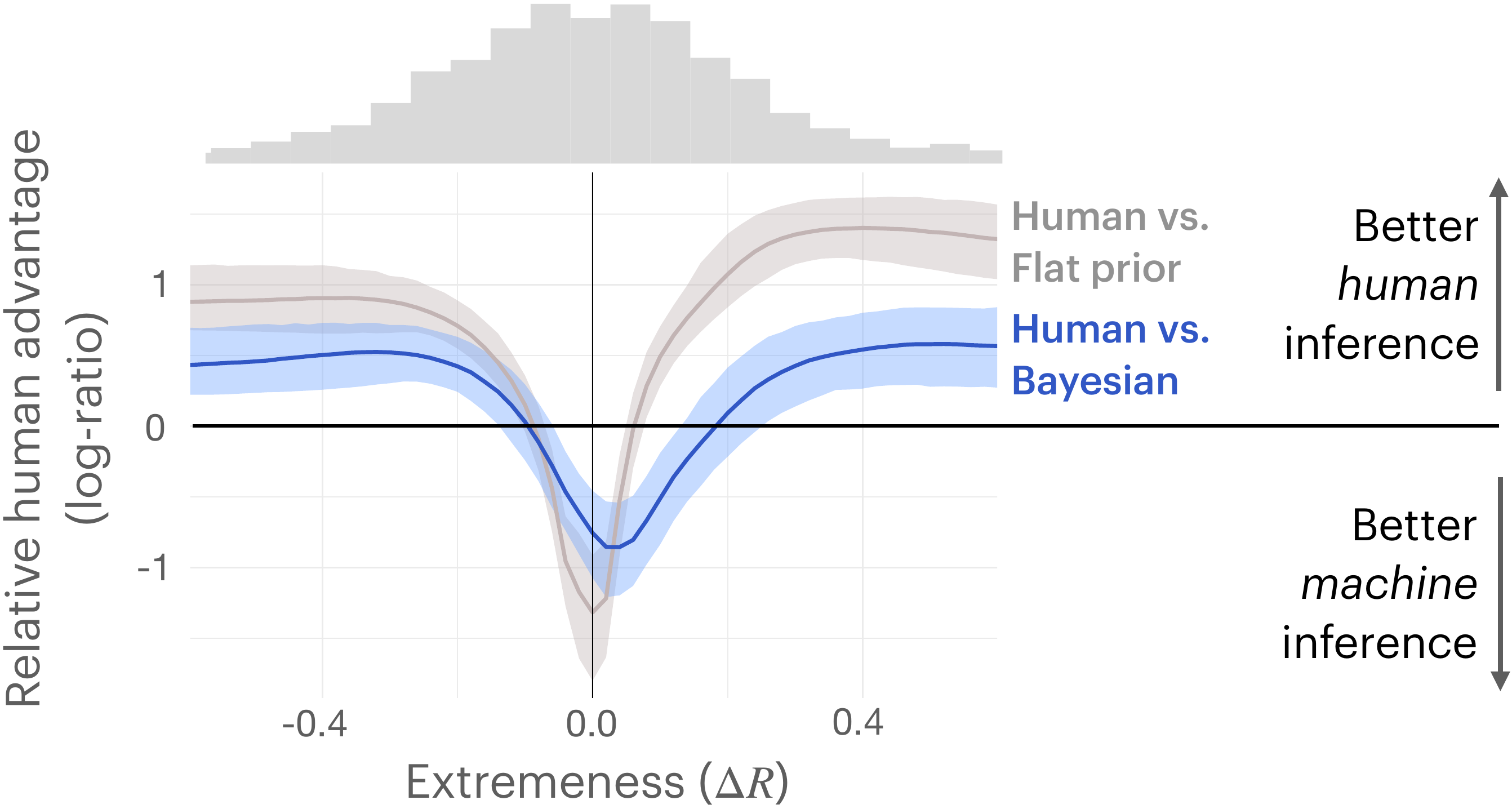}
  \caption{Mean estimated advantage for participants over statistical machines (for inferring the true $\mu$) subject to sample extremeness. Intervals are 95\% CIs, with values above zero indicating a human advantage. The histogram (top) shows the empirically observed $\Delta R$ distribution.}
  \label{fig:exp1_HumanMachine}
  \vspace{-1mm}
\end{figure}

\vspace{.5\baselineskip}\noindent\textbf{Comparing Human vs. Machine Resilience: }
Figure~\ref{fig:exp1_HumanMachine} illustrates the relative robustness of humans against statistical inference. Specifically, we estimate the log-divergence ratio for Bayesian agents over participants: $log(|{\Delta \mu_{Bayes|Flat}}/{\Delta \mu_{Human}}|)$.
An estimate above 0 suggests superior inferential performance for humans compared to statistical machines, while a value below zero indicates the opposite. When $\Delta R=0$ (a sample fully consistent with the ground truth), uninformed Bayesians perform $3.82$ times better than humans (CI: [$2.57, 5.81$]). Similarly, informed Bayesian inference is expected to be $2.13$ times closer to the real data-generating process than the average participant (CI: [$1.59, 2.97$]).

However, as data becomes more extreme, the model predicts an inversion of this relationship. For instance, at $\Delta R = 0.2$ (where 40.7\% of the data has $|\Delta R| \geq 0.2$), humans exhibit similar performance to an informed Bayesian ($1.12\times$ advantage, CI: [$0.81, 1.52$]) while showing a substantial advantage over a flat-prior agent ($2.97\times$, CI: [$2.31, 3.9$]). This advantage widens when $\Delta R = 0.4$ (with 14\% of samples having $|\Delta R| \geq 0.4$). Here, an average human will reliably outperform both informed and uninformed Bayesians by factors of $1.73$ (CI: [$1.33, 2.16$]) and $4.04$ (CI: [$3.24, 5.06$]), respectively.

\subsubsection{How good are the agents at characterizing uncertainty?}

\begin{figure}[t]
\center
\includegraphics[width=1\linewidth]
{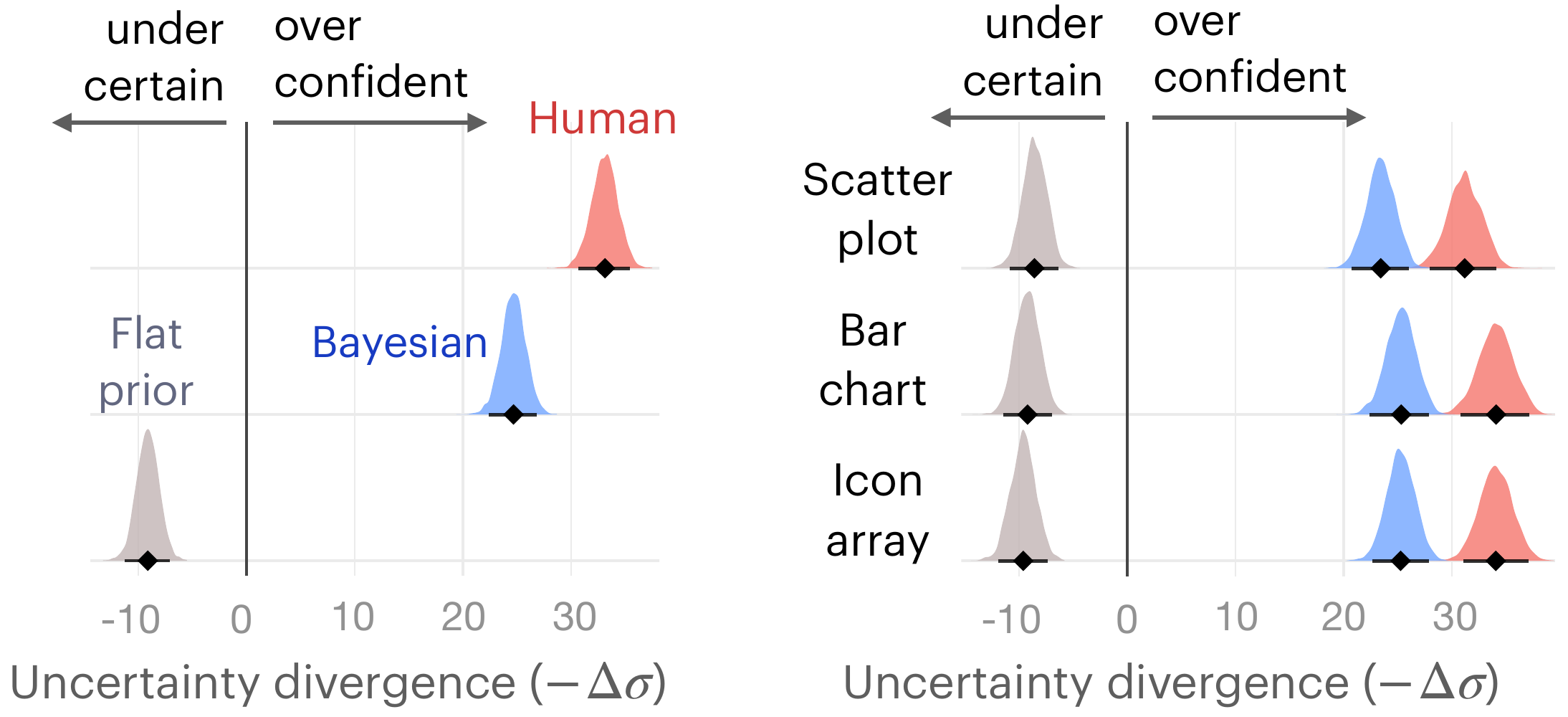}
  \caption{Estimated mean accuracy ($\pm$95\% CIs) in characterizing uncertainty at the aggregate level (left) and by visualization type.
  }
  \label{fig:exp1_sigma}
    \vspace{-4mm}

\end{figure}

Figure \ref{fig:exp1_sigma}-left displays the mean accuracy of three agents in characterizing $\sigma_{Truth}$ (the uncertainty in the ground truth model). The flat prior agent demonstrates a more conservative and accurate inference of true uncertainty ($-9.15$, CI: [$-11.3, -7.12$]). In contrast, both humans and informed Bayesian agents seemed overconfident in their uncertainty estimation ($33.1$, CI: [$30.7, 35.5$] for humans vs. $24.7$, CI: [$22.4, 26.8$] for a Bayesian). The mean accuracy for each visualization type (Figure \ref{fig:exp1_sigma}-right) aligns with the aggregate estimate.

\subsection{Discussion}

Overall, participants were worse at inferring the true slope than at least one of the two Bayesian agents. These results are partially consistent with H1. Surprisingly, however, the flat prior achieved the highest accuracy. This suggests that subject-provided priors were less reliable than the likelihood function implied by the sample. Despite being globally less optimal, participants exhibited better resiliency to spurious samples, especially in high-consensus topics and with smaller sample sizes. Interestingly, a cluster of zero-divergence human responses (Figure~\ref{fig:exp1_lineCharts}, particularly at high consensus) suggests that many participants `correct' for extreme samples, leading to better inference. The model estimates that a typical participant will begin to outperform an informed Bayesian when $|\Delta R| \geq 0.2$ (40.7\% of the data). The flat-prior agent, though best performing in the aggregate, was by comparison highly susceptible to spurious samples. These results support H2.

Surprisingly, participants demonstrated higher inference accuracy with icon arrays than scatterplots, despite the latter's perceptual superiority for bivariate data. This unexpected result (inconsistent with H3) may be attributed to the advantages of icon arrays in conditional probabilistic reasoning~\cite{ottley2015improving}. Such cognitive benefits might outweigh the perceptual advantages of scatterplots. However, it is noteworthy that the spread in performance was most pronounced in icon arrays, indicating a wider variability in response quality, even when this visualization tended to lead to more accurate inference on average. Lastly, and consistent with H4, participants were less accurate in characterizing model uncertainty compared to both Bayesian agents.

\section{Experiment II}

Exp.~1 shows that consensus, a proxy for the level of social agreement,  strongly impacts participants' inferences. This factor could hypothetically influence \emph{when} human analysts might resort to corrective, non-statistically normative heuristics, by drawing upon their socially rooted knowledge. In this experiment, we explore if this social information can compensate for uncertainty in the data-generating process. Specifically, in the presence of heightened statistical uncertainty, we investigate whether participants can leverage their domain intuition to attain a relatively better inference as compared to statistical agents.

\subsection{Hypotheses}

We expect participants to handle increased uncertainty  better than Bayesian agents, particularly when they can compensate with social information:

\vspace{.3\baselineskip}\noindent\textbf{H5 --} The gap between participants and statistical agents will decrease under higher uncertainty. Participants, benefiting from an increased ratio of intuitive, social to statistical information, are expected to reduce the overall performance advantage for Bayesian agents seen in Exp.~1.

\vspace{.3\baselineskip}\noindent\textbf{H6 --} Participants will strategically employ corrective heuristics at higher rates when faced with both high consensus and high uncertainty, making them even more resilient to spurious samples. This behavior will manifest as an interaction between consensus and uncertainty.

\subsection{Participants, Experimental Design, and Procedures}
We recruited 148 participants from Prolific (95 males 50 females 3 others), with a mean age of 37.4 for a \$5 compensation (\$16.6/hour on average). \hl{Self-reported education levels were: 42 high school, 21 associate, 60 bachelor's, 17 master's, 5 doctorate, and 3 unspecified.} The experiment design and procedures were similar to Exp.~1. However, this experiment only employs scatterplots. Participants completed 24 trials. Half the trials were sampled from a model with decreased uncertainty ($\frac{1}{2} \sigma_{Truth}$) and the other half were sampled from a higher-uncertainty model ($2\sigma_{Truth}$). Uncertainty was manipulated as a within-subject factor. We slightly reduced $\sigma_e$ from 0.45 to 0.3 (see Equation~\ref{eq:model}) to allow for more uncertainty control via $\sigma_{Truth}$.

\subsection{Results}

Participants completed the experiment in 18.1 minutes on average (\emph{sd} = 10.4). We analyzed the results by fitting a Bayesian regression model similar to  Exp.~1 (\S\ref{sec:bayes_model}), removing the visualization effect and replacing it with \emph{uncertainty} as a factor (two levels: {low} and {high}). We first investigate how increasing (or decreasing) uncertainty affects inference accuracy ($\Delta \mu$) for the three agents. We then analyze the interaction between social consensus and uncertainty.

\subsubsection{How does uncertainty in the data-generating process impact inference?}

\begin{figure}[]
\center
\includegraphics[width=.95\linewidth]
{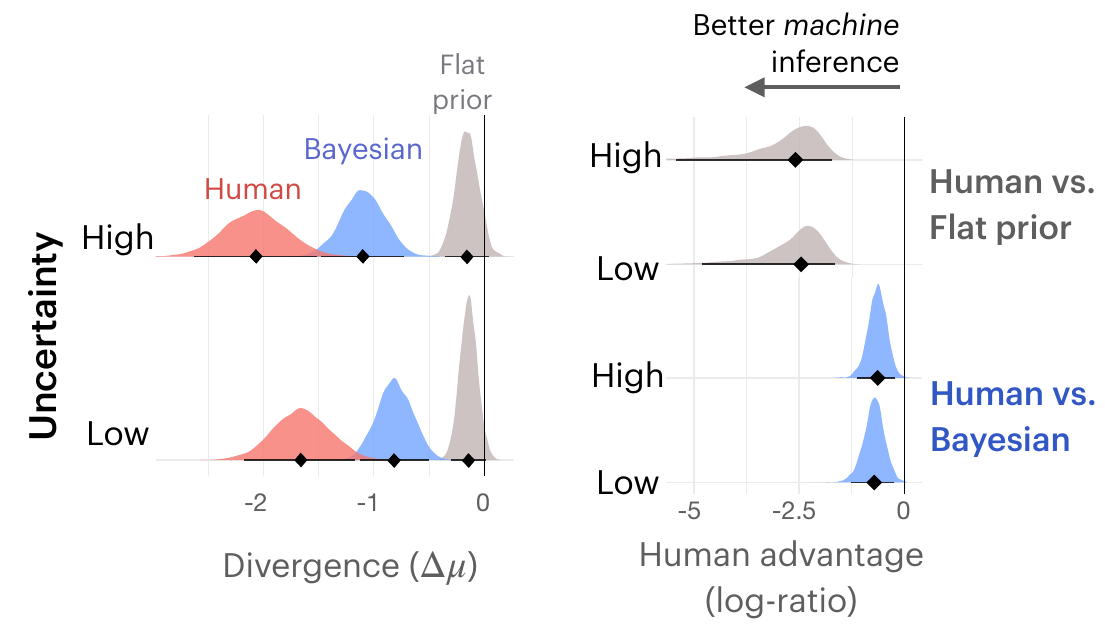}
  \caption{Left: Mean estimated divergence ($\Delta \mu \pm$ 95\% CI). Right: comparison of human and machine accuracy at each uncertainty level.}
  \label{fig:exp2overall}
  \vspace{-4mm}
\end{figure}

Figure~\ref{fig:exp2overall}-left shows the mean accuracy of slope inference for the three agents by uncertainty. 
Informed Bayesian agents and participants do slightly better when uncertainty is low. However, as depicted in Figure~\ref{fig:exp2overall}-right, the gap (represented by the log-ratio) between participants and the two statistical machines remains relatively consistent regardless of the uncertainty level in the data-generating process. Specifically, on average, the flat-prior agent is 11.7 times (CI: [$5.21, 121.51$]  better than humans when the uncertainty is low, and 13.33 times (CI: [5.64, 225.88] better at high uncertainty. The informed Bayesian also exhibits better performance over humans, though at a lesser degree ($2.04\times$ advantage, CI: [$1.29, 3.53$] at low uncertainty vs. $1.89\times$, CI: [$1.26, 3.06$] at high uncertainty. Overall, the observed human-machine gap appears unaffected by changes in uncertainty.

\subsubsection{Can social information compensate for uncertainty?}

\begin{figure}[t]
\center
\includegraphics[width=0.9\linewidth]{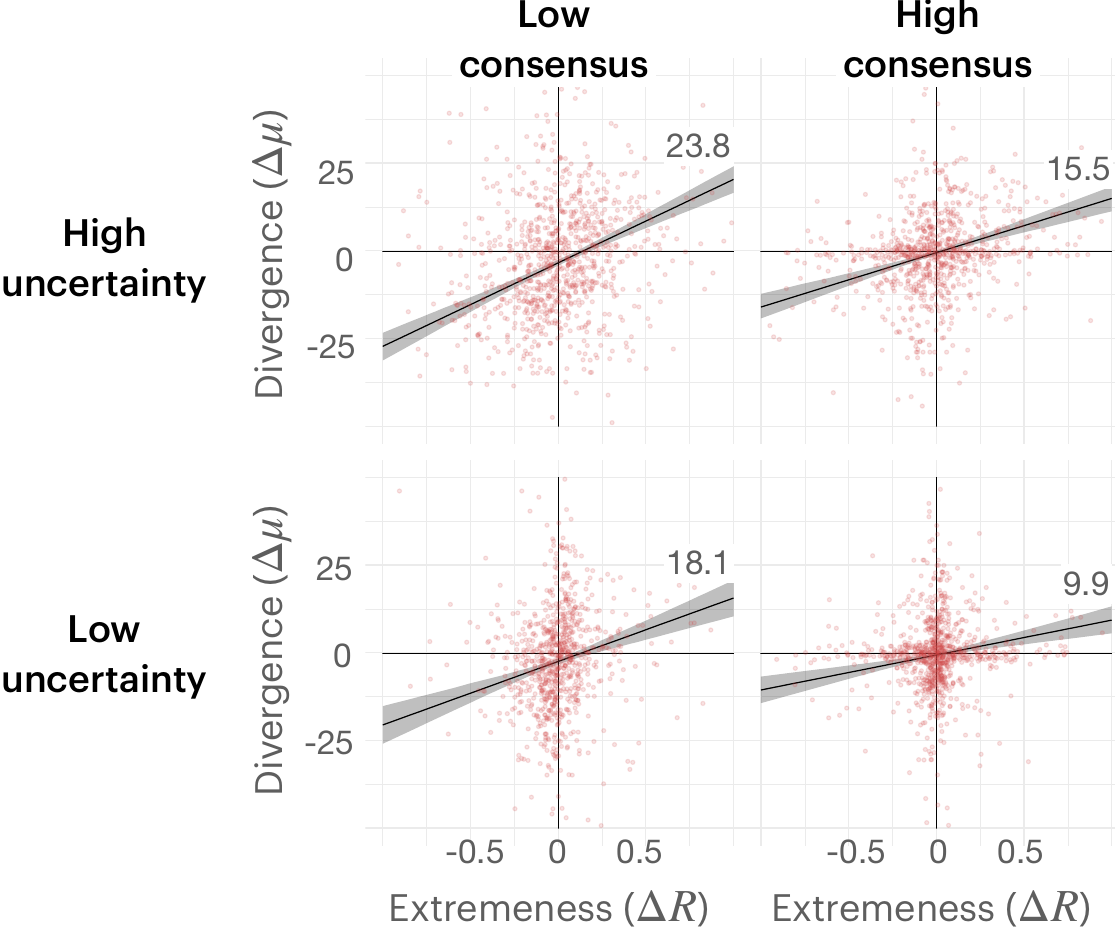}
  \caption{Participants' sensitivity to sample extremeness, showing the interplay between social consensus and uncertainty. Lines show model estimates of $\Delta \mu / \Delta R$ ($\pm$ 95\% CIs). Points correspond to empirical responses. A weaker correlation implies more resilience to spurious data.}
  \label{fig:exp2_lineCharts}
  \vspace{-4mm}
\end{figure}

Figure \ref{fig:exp2_lineCharts}  illustrates how sample extremeness affects participants' inference across different levels of uncertainty and social consensus. In general, heightened uncertainty increases susceptibility to spurious samples, but this sensitivity seems to be moderated by social consensus.
For instance, when uncertainty is high and the consensus is low, the $\Delta \mu / \Delta R$ rate reaches its peak at 23.8 (CI: [$19.96, 27.67$]). However, higher consensus reduces this rate to 15.48 (CI: [$11.76, 18.82$]). Likewise, lower uncertainty decreases the divergence-extremeness rate from 23.8 to 18.07 (CI: [$12.80, 23.33$]). The rate is at its minimum when uncertainty is low, but consensus is high (9.95, CI: [$6.17, 13.80$]).

We investigated the interplay between uncertainty and social consensus using the WAIC criterion~\cite{vehtari2017practical}. A reduced model, which excludes an interaction effect, explained over 99\% of the weights compared to the full model, indicating the absence of an interaction. Accordingly, the observed results can be attributed to a linear combination of social consensus and low uncertainty, both of which appear to enhance participants' ability to disregard spurious data.

\subsection{Discussion}

The results do not support the notion that participants use social information to compensate for statistical uncertainty. Instead, the results suggest a simpler explanation: participants benefit, separately, from increased social consensus and reduced uncertainty in the data-generating model. Consequently, we find no evidence to support an interaction as put forth by H6.  Likewise, there is no indication that participants can narrow the advantage for statistical machines at higher uncertainty levels. On average, human inference performance remains less accurate than that of Bayesians, and this performance gap persists consistently across the two uncertainty levels (contrary to H5). Therefore, while social information may offer inferential benefits, it seems unlikely to aid visualization observers in reducing inherent statistical uncertainties.

\section{General Discussion}

\subsection{Human versus Statistical Inference}

Our findings suggest that human analysts and Bayesian machines excel in different scenarios. While Bayesian agents generally provide more optimal inference, humans showed less deviation from the ground truth with spurious data. For instance, at the 60th percentile of extreme data, humans matched informed Bayesians and outperformed flat prior agents by a factor of 3. This advantage increased at the 86th percentile, with factors of 1.7 and 4, respectively. Participants appear to rely on their internal models to disregard extreme samples, reducing bias. Even when the Bayesian agent is provided with a human prior model, the human inference remains less affected by extreme data.

\hl{These findings support our central hypothesis that analysts' intuitive and non-rational heuristics, such as ignoring or underweighting seemingly implausible data samples, can be advantageous. While overlooking improbable evidence can generally be problematic}~\cite{koehler2004thinking,camerer1989decision}, \hl{this tendency may prove useful in exploratory data analysis, by making analysts more skeptical of patterns they encounter in visualizations, effectively reducing the rate of false discoveries from visualizations. For instance, when viewing a visualization suggesting a positive increase in sales that the analyst believes to be counterintuitive, the analyst may be justified in underweighting this evidence and concluding that there was no substantial increase in this case. Although such a conclusion may be non-statistically rational given the data at hand, it could prevent the analyst from committing a false discovery.}

\hl{Prior work by Zgraggen et al. suggests that visual analysts frequently make Type I errors, such as concluding that there is an effect, like a positive bivariate correlation or a non-zero difference between two groups, when in reality there are none}~\cite{zgraggen2018investigating}. In contrast, our findings suggest that, on average, individuals may be more adept at discerning true effects from noise by using their internal models and heuristics to resist spurious patterns. This ability, however, appears contingent on having good intuition about the underlying phenomena. \hl{Moreover, our results indicate that formalizing this intuitive process, for example, by enforcing or encouraging universally rational (e.g., Bayesian) inference, could be counterproductive. Visual analytics systems should retain a role for analyst intuition, even when their conclusions may seem biased.}

Despite the advantage at interpreting extreme data samples, human inference seems to exhibit higher variability. Individual analyst responses may thus be further from the truth, even though collectively, humans may be less prone to overinterpreting extreme samples. This variability may be attributed to noise from two sources: the perception of visualizations, and the setting of priors and posteriors through graphical elicitation devices. These channels allow for visual-perceptual, cognitive, and manual errors to seep in, thereby affecting response quality. Computational statistical agents, on the other hand, are unaffected by these noise sources, and may thus provide a less varied response. Similarly, in terms of inferring uncertainty, participants consistently exhibited overconfidence. The Bayesian agent's response, while also expressing higher certainty about the data-generating process than warranted, was closer to the ground truth. Conversely, the uninformed agent provided a more conservative assessment that was almost always closer to the true uncertainty in the generating model.

\subsection{People-Machine Collaboration for Inferential Analysis}
Our findings suggest the potential for collaboration between humans and statistical (or AI-based) agents for inference-making, enabling each to complement the other's limitations~\cite{wang2020human, steyvers2022bayesian}. This collaboration could take the form of delegation~\cite{mozannar2022teaching, lai2022human, bondi2022role, hemmer2023human}, where each agent (human or statistical) takes responsibility for certain inferential tasks depending on the sample size and the quality of the human intuition. For example, systems can delegate the responsibility of making inferences from large datasets to a rational statistical agent. Conversely, humans could manually handle interpretive scenarios involving smaller and less ideal datasets, where analyst intuition are likely to provide an edge. One challenge to enabling this sort of collaborative inference-making is to predict the potential extremeness of the data at hand so that effective delegation can take place~\cite{fugener2022cognitive, pinski2023ai}. A second challenge is the need to reduce variability in individual human responses without compromising their advantages. We suspect that improving knowledge elicitation in visualizations (e.g., by redesigning and validating graphical model representation techniques~\cite{karduni2020bayesian}) could help reduce response variability by reducing noise in externalizing human priors and posteriors.
The idea of using AI for certain aspects of the data analysis pipeline, such as automating the creation of machine learning models~\cite{drozdal2020trust, wang2021autods}, has indeed been suggested. However, data professionals continue to express reservations about this model~\cite{wang2019human}. By contrast, a collaborative workflow that genuinely involves analysts could garner more positive reception over `Auto Data Science' approaches~\cite{karmaker2021automl,aggarwal2019can}.

\subsection{The Utility of a Biased Mind}

In visual analytics, the concept of mixed-initiative collaboration between humans on one hand and algorithms and statistical models on the other is well established~\cite{endert2014human,cui2019visual}. Recent efforts in this space aim to algorithmically identify deviations from statistical rationality~\cite{kim2019bayesian,karduni2020bayesian}, and introduce interventions to mitigate biases in human decision-making~\cite{cho2017anchoring,wall2017warning,bedek2018methods,dimara2018task}. This view often holds that any deviation from idealized, analytic normativeness is problematic and should be reduced to a minimum. Our findings, however, suggest that analyst deviations in graphical inference could indeed be useful. Thus, rather than solely focusing on `debiasing' human reasoning (e.g., by seeking to align human responses with rational models~\cite{wu2023rational}), our work suggests the need for a more balanced perspective that acknowledges the utility of both rational and intuitive thinking~\cite{dane2012should,sadler2004intuitive,haselton2009adaptive,gigerenzer2009homo}, even if the latter is subject to biases. Instead of attempting to eliminate these biases in visual analytics, a more fruitful approach involves considering how to productively leverage analyst intuition. However, it is important to acknowledge that not all biases are useful, and some may lead to systematic errors and bad judgments, especially in response to perverse incentives (e.g., \hl{the pressure to infer a positive (non-null) conclusion from data for publication purposes}~\cite{lindner2018scientific,forstmeier2017detecting,nosek2012scientific}). With this understanding, systems can learn to trust human judgment and, conversely, to intervene with computational support when necessary, for example, by furnishing a (more) rational solution for human consideration.
\section{Limitations and Future Work}

There are limitations to consider when interpreting our results.  First, our experiments utilized varying sample sizes, ranging from `small' to `large'. Yet, even the largest size ($N=30$) presents a relatively small amount of data. Larger datasets, in particular, could lead to a greater advantage for statistical machines over humans than we observed in our study. Future studies should therefore include a wider range of sample sizes, including more realistically large datasets. Second, the ground-truth models we used were primarily based on common-sense semantics. These models do not necessarily reflect more contentious or expertise-dependent topics. \hl{Additionally, our study tested a limited range of visualizations (scatterplots, bar charts, and icon arrays), hence the results may not generalize to other representations. The involvement of crowdsourced participants, likely non-experts, is another limitation.} Lastly, the workflow in our experiments was highly controlled. In reality, analysts are more likely to engage in fluid, open-ended visual exploration. Unlike participants, analysts also have the autonomy to choose which visualizations to view based on their priors. Future studies should therefore attempt to replicate our findings with practicing data scientists under more realistic tool usage conditions.
\section{Conclusion}

We investigated human inference-making from (at times noisy) visualizations, contrasting the accuracy of human judgments with those of Bayesian agents. Our central hypothesis was that, while humans exhibit non-rational tendencies, they may possess advantages in certain scenarios. The experiment findings support this notion: Although participants were generally less optimal than Bayesian benchmarks, they surpassed machines in specific instances and were better at discerning (and disregarding) spurious datasets. However, our results also indicate that humans exhibit higher variability and consistently exhibit overconfidence. Our findings suggest future design avenues where intuitive thinking on the part of analysts and statistical models can complement each other. Furthermore, our work challenges the assumption that normative rationality is always ideal and shows that analyst gut reactions could provide an advantage in certain visual analytic contexts.


\acknowledgments{This research was supported by the National Science Foundation under award 1942429, and by the Office of Science, US Department of Energy under contract DE-AC02-06CH11357. KR was also supported by the Argonne faculty sabbatical program.}

\bibliographystyle{abbrv-doi-hyperref}

\interlinepenalty=10000
\balance

\bibliography{00_template}

\appendix 

\end{document}